\documentclass[%
reprint,
superscriptaddress,
showpacs,
amsmath,amssymb,
prc,
floatfix, ]%
{revtex4-1}

\usepackage{color}

\usepackage{graphicx}
\usepackage{dcolumn}
\usepackage{bm}
\usepackage[dvipdfmx,bookmarks=true,colorlinks,%
            citecolor=blue,linkcolor=blue,anchorcolor=blue,filecolor=blue,urlcolor=blue,%
           ]{hyperref}

\begin{document}


\title{Nonaxial-octupole $Y_{32}$ correlations in $N = 150$ isotones
       from multidimensional constrained covariant density functional theories}

\author{Jie Zhao}
 \affiliation{State Key Laboratory of Theoretical Physics,
              Institute of Theoretical Physics, Chinese Academy of Sciences, Beijing 100190, China}
\author{Bing-Nan Lu}
 \affiliation{Physics Department, Faculty of Science, University of Zagreb, Zagreb 10000, Croatia}
 \affiliation{State Key Laboratory of Theoretical Physics,
              Institute of Theoretical Physics, Chinese Academy of Sciences, Beijing 100190, China}

\author{En-Guang Zhao}
 \affiliation{State Key Laboratory of Theoretical Physics,
              Institute of Theoretical Physics, Chinese Academy of Sciences, Beijing 100190, China}
 \affiliation{Center of Theoretical Nuclear Physics, National Laboratory
              of Heavy Ion Accelerator, Lanzhou 730000, China}
\author{Shan-Gui Zhou}
 \email{sgzhou@itp.ac.cn}
 \affiliation{State Key Laboratory of Theoretical Physics,
              Institute of Theoretical Physics, Chinese Academy of Sciences, Beijing 100190, China}
 \affiliation{Center of Theoretical Nuclear Physics, National Laboratory
              of Heavy Ion Accelerator, Lanzhou 730000, China}

\date{\today}

\begin{abstract}
The non-axial reflection-asymmetric $\beta_{32}$ shape in some transfermium
nuclei with $N=150$, namely $^{246}$Cm, $^{248}$Cf, $^{250}$Fm, and $^{252}$No
are investigated with multidimensional constrained covariant density functional theories.
By using the density-dependent point coupling covariant density functional theory 
with the parameter set DD-PC1 in the particle-hole channel, it is found that, 
for the ground states of $^{248}$Cf and  $^{250}$Fm, the non-axial octupole deformation
parameter $\beta_{32} > 0.03$ and the energy gain due to the $\beta_{32}$
distortion is larger than 300 keV.
In $^{246}$Cm and $^{252}$No, shallow $\beta_{32}$ minima are found.
The occurrence of the non-axial octupole $\beta_{32}$ correlations is mainly from a pair of
neutron orbitals $[734]9/2$ ($\nu j_{15/2}$) and $[622]5/2$
($\nu g_{9/2}$) which are close to the neutron Fermi surface and a pair of
proton orbitals $[521]3/2$ ($\pi f_{7/2}$) and $[633]7/2$ ($\pi i_{13/2}$)
which are close to the proton Fermi surface.
The dependence of the non-axial octupole effects on the form of energy
density functional and on the parameter set is also studied.
\end{abstract}


\maketitle


The majority of observed nuclear shapes is of spheroidal form which
is usually described by $\beta_{20}$.
The existence of the nonaxial-quadrupole (triaxial) deformation $\beta_{22}$
~\cite{Starosta2001_PRL86-971, Odegard2001_PRL86-5866, Meng2010_JPG37-064025}
and the axial octupole deformation $\beta_{30}$~\cite{Butler1996_RMP68-349}
in atomic nuclei have also been confirmed both experimentally and theoretically.
However, there is no a priori reason to neglect the nonaxial-octupole deformations,
especially the $\beta_{32}$ deformation~\cite{Hamamoto1991_ZPD21-163,
Skalski1991_PRC43-140, Li1994_PRC49-R1250}.
The pure $\beta_{32}$ deformation has a tetrahedral symmetry with a symmetry group $T_d^D$.
The existence of non-trivial irreducible representations of
this group makes it possible for a nuclei to have large energy gaps
in their single-particle levels, thus increasing it's stability~\cite{Dudek2010_JPG37-064032}.
It has been anticipated that $\beta_{32}$ deformation occurs in the ground states of
some nuclei with special combinations of the neutron and
proton numbers~\cite{Li1994_PRC49-R1250, Dudek2010_JPG37-064032, Dudek2002_PRL88-252502}.
Recently, lots of theoretical studies focus on this nuclear shape,
either from the $T_d^D$-symmetric single particle spectra~\cite{Li1994_PRC49-R1250,
Dudek2002_PRL88-252502,Dudek2003_APPB34-2491,Dudek2007_IJMPE16-516} or
from various nuclear models including the macroscopic-microscopic
model~\cite{Dudek2002_PRL88-252502, Dudek2007_IJMPE16-516, Schunck2004_PRC69-061305R,
Dudek2006_PRL97-072501}, the Skyrme Hartree-Fock (SHF), SHF+BCS, or Skyrme Hartree-Fock-Bogoliubov
models~\cite{Dudek2007_IJMPE16-516, Schunck2004_PRC69-061305R,
Yamagami2001_NPA693-579, Olbratowski2006_IJMPE15-333, Dudek2006_PRL97-072501,
Zberecki2009_PRC79-014319, Takami1998_PLB431-242, Zberecki2006_PRC74-051302R},
and the Reflection Asymmetric Shell Model (RASM)~\cite{Gao2004_CPL21-806, Chen2008_PRC77-061305R}.
In particular, Dudek et al. predicted that a negative-parity band in $^{156}$Gd
is a favorable candidate to manifest tetrahedral symmetry~\cite{Dudek2006_PRL97-072501}
which has stimulated several interesting experimental studies~\cite{Bark2010_PRL104-022501,Jentschel2010_PRL104-222502}.

Nowadays the study of nuclei with $Z\sim 100$ becomes more and more important
because it not only reveals the structure of these nuclei themselves
but also provides significant information for superheavy nuclei~\cite{Herzberg2006_Nature442-896,
Zhang2011_PRC83-011304R, Zhang2012_PRC85-014324, Zhang2012_arxiv1208.1156}.
One of the relevant and interesting topics is how to explain the low-lying
$2^-$ states in some $N = 150$ even-even nuclei.
In these nuclei, the bandhead energy $E(2^-)$ of the lowest $2^-$ bands
is very low~\cite{Robinson2008_PRC78-034308}.
It is well accepted that the octupole correlation is responsible for it.
For example, a quasiparticle phonon model with octupole correlations included was
used to explain the excitation energy of the $2^-$ state of the
isotones with $N = 150$ and the octupole correlation is mainly due to
the neutron two-quasiparticle configuration $9/2^-[724] \otimes 5/2^+[622]$
and proton two-quasiparticle configurations $9/2^+[633] \otimes 5/2^-[521]$
or $7/2^+[633] \otimes 3/2^-[521]$~\cite{Jolos2011_JPG38-115103}.
In all these configurations, the third components of the angular momenta
of like-quasiparticles $K$ differ by 2, i.e.,
$Y_{32}$-correlation should play an important role.
In Ref.~\cite{Chen2008_PRC77-061305R}, Chen et al. proposed that the non-axial
octupole $Y_{32}$-correlation results in the experimentally observed low-energy $2^-$
bands in the $N = 150$ isotones and the RASM calculations
reproduces well the experimental observables of these $2^-$ bands.
It was also predicted that the strong nonaxial-octupole effect may persist
up to the element 108~\cite{Chen2008_PRC77-061305R} and
play a crucial role in determining the shell stability in
even heavier nuclei~\cite{Chen2012_in-prep}.
Pronounced shell effects were also found for $N=16$, 40,
and 110 when combining nonaxial octupole deformations~\cite{Heiss1999_PRC60-034303}.

In this Brief Report we present a microscopic and self-consistent study of
the $Y_{32}$ effects in the $N = 150$ isotones
under the framework of the covariant density functional
theory (CDFT)~\cite{Serot1986_ANP16-1, Reinhard1989_RPP52-439, Ring1996_PPNP37-193,
Bender2003_RMP75-121, Vretenar2005_PR409-101, Meng2006_PPNP57-470}.
We use the recently developed multi-dimensional constraint (MDC)
CDFT~\cite{Lu2012_PRC85-011301R, Lu2012_in-prep}
in which not only the axial~\cite{Meng2006_PRC73-037303, Lu2011_PRC84-014328}
but also the reflection symmetries~\cite{Geng2007_CPL24-1865} are broken.
For the parametrization of the nuclear shape, the conventional ansatz in
mean-field calculations
\begin{equation}
 \beta_{\lambda\mu} = {4\pi \over 3AR^\lambda} \langle Q_{\lambda\mu} \rangle,
 \label{eq:01}
\end{equation}
is adopted, where $Q_{\lambda\mu} = r^\lambda Y_{\lambda \mu} $ 
is the mass multipole operator.
The nuclear shape is assumed to be invariant under the reversion of $x$ and $y$ axes
in MDC-CDFT, i.e., the intrinsic symmetry group is $V_{4}$ and all shape
degrees of freedom $\beta_{\lambda\mu}$ deformations with even $\mu$ are allowed.
The CDFT functional can be one of the following four forms:
the meson exchange or point-coupling nucleon interactions combined with
the non-linear or density-dependent couplings~\cite{Lu2012_PRC85-011301R, Lu2012_in-prep}.
In the present work, the pairing effect is taken into account by using
the BCS approximation with a separable pairing
force~\cite{Tian2009_PLB676-44, Tian2009_PRC80-024313, Niksic2010_PRC81-054318}.

\begin{table}
\caption{\label{tab:DDPC1} %
The quadrupole deformation $\beta_{20}$, non-axial octupole deformation
$\beta_{32}$, and hexadecapole deformation $\beta_{40}$ together with
binding energies $E_\mathrm{cal}$ for the ground states of
$N=150$ nuclei calculated with DD-PC1. $E_\mathrm{depth}$ denotes the energy
difference between the ground state and the point with $\beta_{32} = 0$
in the potential energy curves shown in Fig.~\ref{fig:b32}.
$E_\mathrm{exp}$ denotes experimental binding energies taken
from~\cite{Audi2011_Private}. All energies are in MeV.
}
\begin{ruledtabular}
\begin{tabular}{ccccccc}
 Nucleus   & $\beta_{20}$ & $\beta_{32}$ & $\beta_{40}$ & $E_\mathrm{cal}$ & $E_\mathrm{depth}$ & $E_\mathrm{exp}$ \\ \hline
$^{246}$Cm &  0.296       &  0.020       &  0.126       & $-1847.932$ &  0.034          & $-1847.819$ \\
$^{248}$Cf &  0.299       &  0.037       &  0.111       & $-1857.853$ &  0.351          & $-1857.776$ \\
$^{250}$Fm &  0.293       &  0.034       &  0.097       & $-1865.843$ &  0.328          & $-1865.520$ \\
$^{252}$No &  0.293       &  0.025       &  0.083       & $-1872.133$ &  0.104          & $-1871.305$ \\
\end{tabular}
\end{ruledtabular}
\end{table}

First we study the ground states of nuclei $^{246}$Cm, $^{248}$Cf,
$^{250}$Fm, and $^{252}$No by using the density-dependent point coupling
covariant density functional theory with the parameter set
DD-PC1 in the particle-hole channel~\cite{Niksic2008_PRC78-034318}.
The quadrupole deformations $\beta_{20}$, non-axial octupole deformation
$\beta_{32}$, and hexadecapole deformations $\beta_{40}$ together with
binding energies $E_\mathrm{cal}$ are listed in Table~\ref{tab:DDPC1}.
The non-axial quadrupole and axial octupole deformation parameters
are all zero for these nuclei.
As is seen in Table~\ref{tab:DDPC1}, the calculated binding energies
agree very well with the data.
All of these four nuclei are well deformed with the quadrupole deformation parameter
$\beta_{20} \approx 0.3$ and the hexadecapole deformations $\beta_{40} \approx 0.1$.
Superposed with large quadrupole and hexadecapole deformations, finite
values are obtained for the non-axial octupole deformation $\beta_{32}$ for these
$N=150$ isotones.

\begin{figure}
  \includegraphics[width=0.45\textwidth]{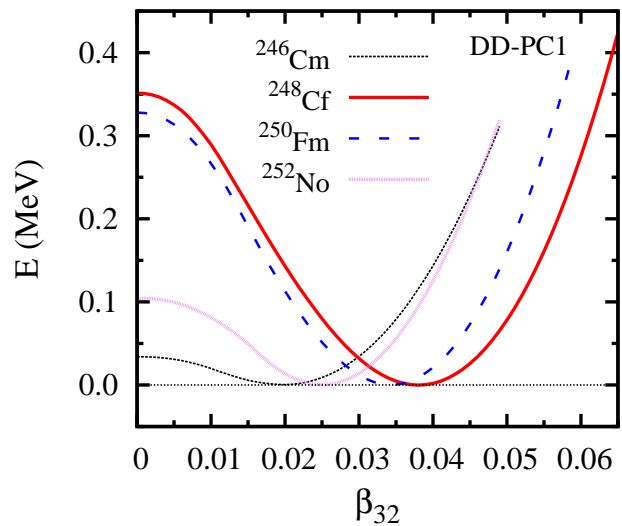}
\caption{\label{fig:b32}(Color online) %
The binding energy $E$ (relative to the ground state) for $N=150$ isotones $^{246}$Cm
(dashed line), $^{248}$Cf (solid line), $^{250}$Fm (long-dashed line), 
and $^{252}$No (dotted line) as a function of the non-axial octupole 
deformation parameter $\beta_{32}$.
}
\end{figure}

In order to see more clearly the development of non-axial octupole
$\beta_{32}$ deformation along the $N=150$ isotonic chain, we preform
one-dimensional constrained calculations and obtain potential energy curves,
i.e., the total binding energy as a function of $\beta_{32}$.
At each point of a potential energy curve, the energy is minimized
automatically with respect to other shape degrees
of freedom such as $\beta_{20}$, $\beta_{22}$, $\beta_{30}$, and $\beta_{40}$, etc.
In Fig.~\ref{fig:b32}, we show potential energy curves for these $N=150$ isotones.
For $^{246}$Cm, the ground state deformation $\beta_{32}=0.020$
(see Fig.~\ref{fig:b32} as well as Table~\ref{tab:DDPC1}).
The potential energy curve is rather flat around the minimum.
We denote the energy difference between the ground state and the point with
$\beta_{32}=0$ by $E_\mathrm{depth}$ which 
measures the energy gain with respect to the $\beta_{32}$ distortion.
For $^{246}$Cm, $E_\mathrm{depth}$ is only $34$ keV.
For $^{248}$Cf, $^{250}$Fm and $^{252}$No, the minima locate at
$\beta_{32} = 0.037$, $0.034$, and $0.025$, respectively.
The corresponding energy gain $E_\mathrm{depth} = 0.351$, $0.328$, and $0.104$ MeV.
It may be hard to conclude that these nuclei have static non-axial
octupole deformations from these results because the potential energy curve
is flat around the minimum and $E_\mathrm{depth}$ is small.
However, the present calculations at least indicate a strong $Y_{32}$-correlation in these nuclei.
Both the non-axial octupole parameter $\beta_{32}$ and the energy gain $E_\mathrm{depth}$
reach maximal values at $^{248}$Cf in the four nuclei along the $N=150$ isotonic chain.
This is consistent with the analysis given in Refs.~\cite{Chen2008_PRC77-061305R,
Jolos2011_JPG38-115103} and the experimental observation that in $^{248}$Cf,
the $2^-$ state is the lowest among these nuclei.

\begin{figure}
 \includegraphics[width=0.48\textwidth]{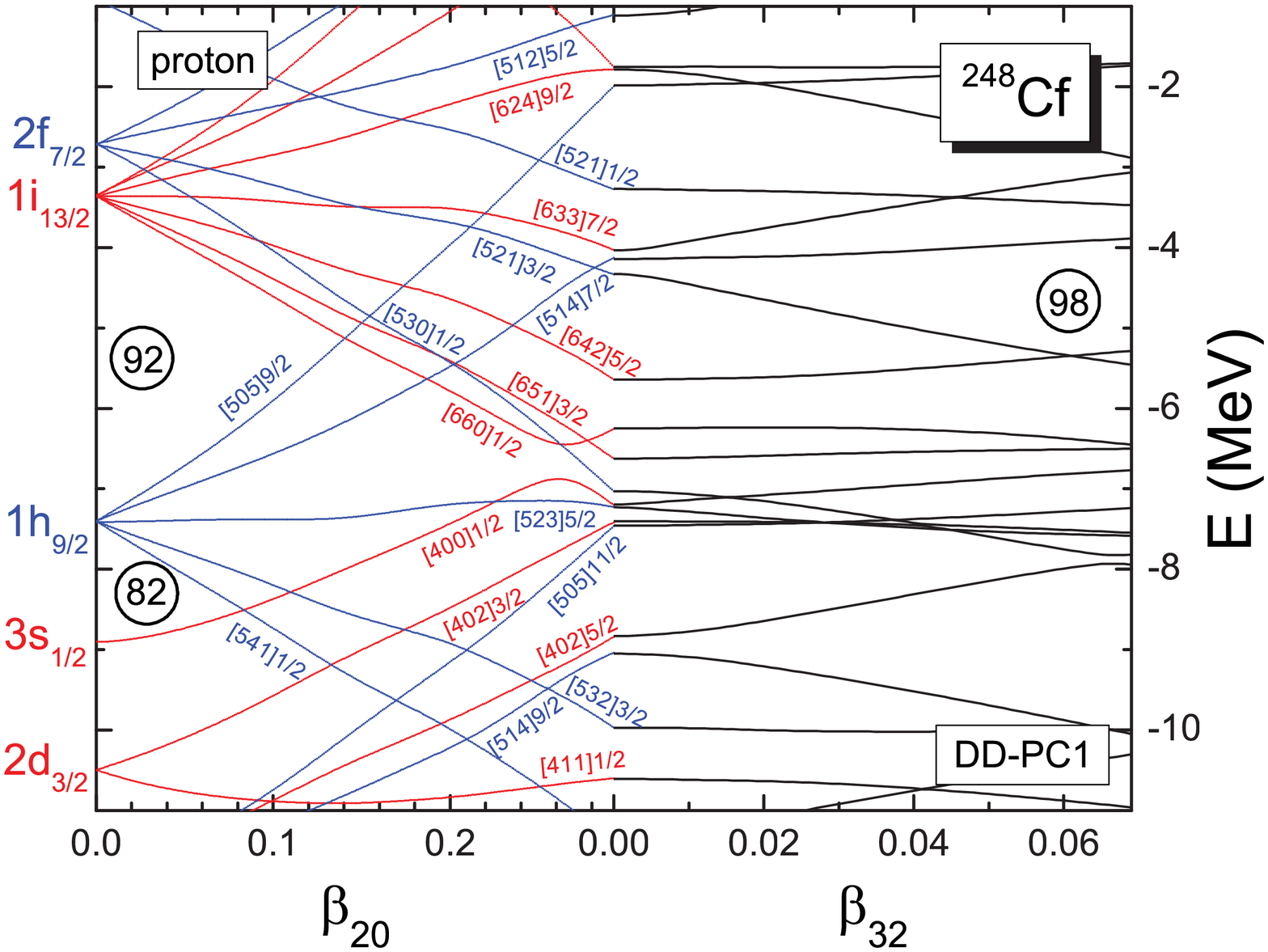}
 \includegraphics[width=0.48\textwidth]{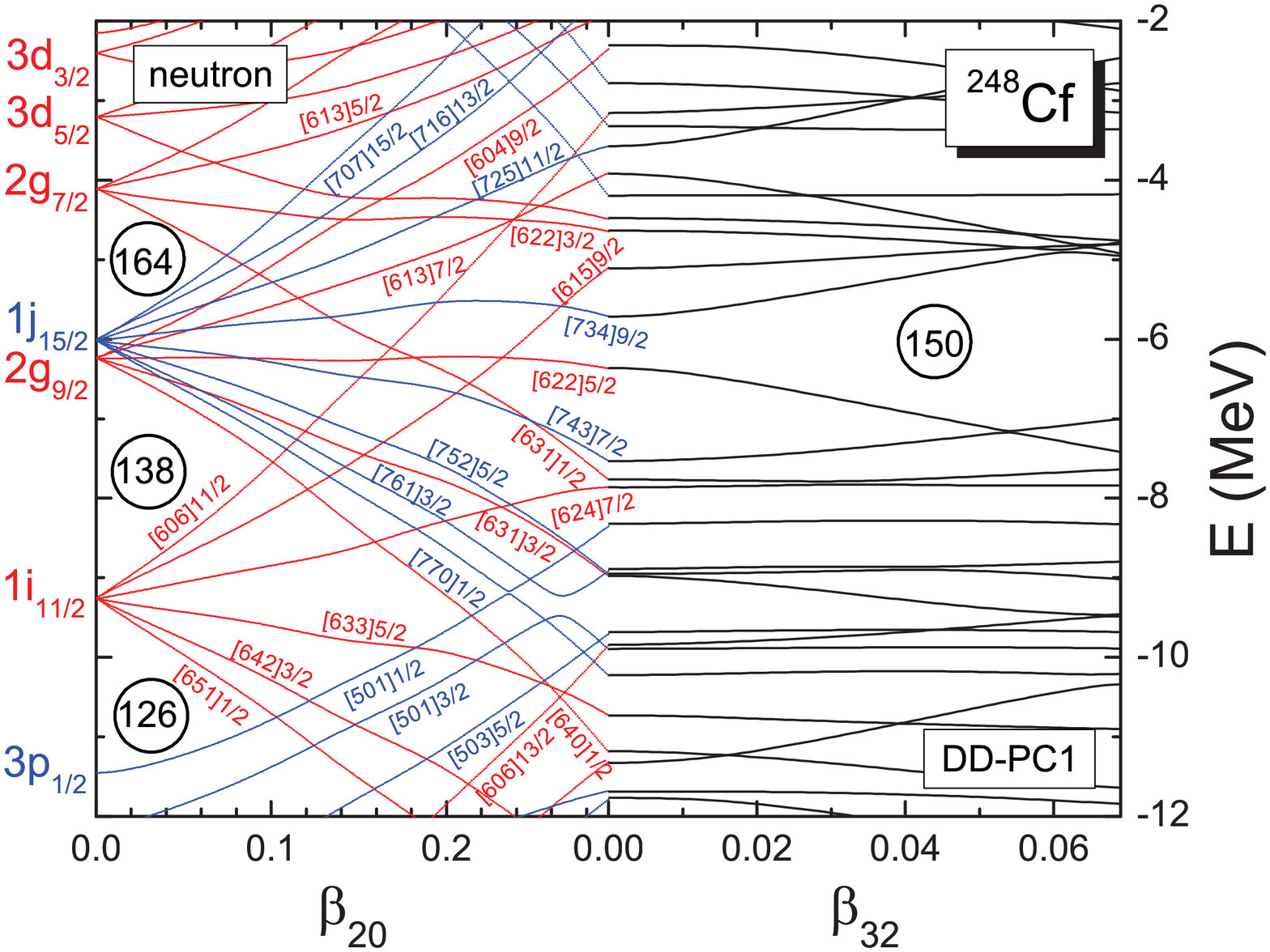}
\caption{\label{fig:lev}(Color online) %
The single-particle levels near the Fermi surface for (a) protons
and (b) neutrons of $^{248}$Cf as functions of quadrupole deformation
$\beta_{20}$ (left side) and of $\beta_{32}$ with $\beta_{20}$ fixed at 0.3 (right side).
}
\end{figure}

The triaxial octupole $Y_{32}$ effects stem from the coupling between pairs of
single-particle orbits with $\Delta j = \Delta l = 3$ and $\Delta K=2$
where $j$ and $l$ are the total and orbit angular momentum of single particles
respectively and $K$ the projection of $j$ on the $z$ axis.
If the Fermi surface of a nucleus lies close to a pair of such orbitals
and these two orbitals are nearly degenerate,
a strong non-axial octupole $\beta_{32}$ effect is expected.

In Fig.~\ref{fig:lev}, we show the proton and neutron single-particle levels
near the Fermi surface for $^{248}$Cf as functions of quadrupole deformation
$\beta_{20}$ on the left side and of $\beta_{32}$ with $\beta_{20}$ fixed at 0.3
on the right side.
In Fig.~\ref{fig:lev}(a), one finds a strong spherical shell closure at $Z=92$.
As discussed in Ref.~\cite{Geng2006_CPL23-1139}, this shell closure
is a spurious one and it is commonly predicted in
relativistic mean field calculations.
The spherical proton orbitals $\pi 2f_{7/2}$ and $\pi 1i_{13/2}$ are very
close to each other and this near degeneracy results in octupole correlations.
The two proton levels, $[521]3/2$ originating from $2f_{7/2}$ and $[633]7/2$
originating from $1i_{13/2}$, satisfying the $\Delta j = \Delta l = 3$ and
$\Delta K=2$ condition, are very close to each other when $\beta_{20}$
varies from 0 to 0.3.
Therefore the non-axial octupole $Y_{32}$ develops and with $\beta_{32}$
increasing from zero, an energy gap appears at $Z=98$.
The appearance of the spurious shell closure at $Z=92$ is mainly due to
the lowering of $\pi 1h_{9/2}$ orbital which also results in that the
proton $[514]7/2$ level lies in between $[521]3/2$ and $[633]7/2$,
thus making the gap at $Z=98$ smaller and weakening the $Y_{32}$ correlation.
Similarly, the spherical neutron orbitals $\nu 2g_{9/2}$ and $\nu 1j_{15/2}$
are very close to each other.
The two neutron levels, $[734]9/2$ originating from $1j_{15/2}$ and $[622]5/2$
originating from $2g_{9/2}$, are also close to each other and they just
lie above and below the Fermi surface.
This leads to the development of a gap at $N=150$ with $\beta_{32}$ increasing.
Therefore it is clear that the $Y_{32}$ correlation in $N=150$ isotones
is from both protons and neutrons and for $^{248}$Cf the correlation
is the most pronounced.
The gap around $N=150$ is larger than that around $Z=98$,
which may indicate that the non-axial octupole effect originating from
neutrons is larger than that from protons.
If there is not a spurious shell gap at $Z=92$,
one can expect more pronounced non-axial octupole correlations from protons.

\begin{table}
\caption{\label{tab:parameter} %
The quadrupole deformation $\beta_{20}$,
octupole deformation $\beta_{32}$, hexadecapole deformation
$\beta_{40}$ together with binding energies $E$ for the ground states of
$N=150$ nuclei calculated with parameter sets PC-PK1, DD-ME1, DD-ME2 and DD-PC1.
$E_\mathrm{depth}$ denotes the energy difference between the ground states and the
point constrained $\beta_{32}$ to zero.
All energies are in MeV.
}
\begin{ruledtabular}
\begin{tabular}{ccccccccc}
 Nucleus    & Parameters & $\beta_{20}$ & $\beta_{32}$ & $\beta_{40}$ & $E$ & $E_\mathrm{depth}$ \\
\hline
$^{246}$Cm  & PC-PK1     & $0.304$      & $0.0$        & $0.129$      & $-1846.273$ & ---     \\
            & DD-ME1     & $0.297$      & $0.0$        & $0.126$      & $-1847.205$ & ---     \\
            & DD-ME2     & $0.296$      & $0.0$        & $0.125$      & $-1846.662$ & ---     \\
            & DD-PC1     & $0.296$      & $0.020$      & $0.126$      & $-1847.932$ & $0.034$ \\
\hline
$^{248}$Cf  & PC-PK1     & $0.306$      & $0.031$      & $0.113$      & $-1857.327$ & $0.139$ \\
            & DD-ME1     & $0.299$      & $0.035$      & $0.107$      & $-1857.718$ & $0.278$ \\
            & DD-ME2     & $0.300$      & $0.039$      & $0.105$      & $-1857.088$ & $0.474$ \\
            & DD-PC1     & $0.299$      & $0.037$      & $0.111$      & $-1857.853$ & $0.351$ \\
\hline
$^{250}$Fm  & PC-PK1     & $0.302$      & $0.029$      & $0.098$      & $-1865.301$ & $0.128$ \\
            & DD-ME1     & $0.293$      & $0.033$      & $0.093$      & $-1866.425$ & $0.305$ \\
            & DD-ME2     & $0.292$      & $0.034$      & $0.091$      & $-1865.505$ & $0.416$ \\
            & DD-PC1     & $0.293$      & $0.034$      & $0.097$      & $-1865.843$ & $0.328$ \\
\hline
$^{252}$No  & PC-PK1     & $0.300$      & $0.013$      & $0.085$      & $-1870.414$ & $0.001$ \\
            & DD-ME1     & $0.293$      & $0.023$      & $0.080$      & $-1873.412$ & $0.089$ \\
            & DD-ME2     & $0.292$      & $0.024$      & $0.078$      & $-1872.235$ & $0.125$ \\
            & DD-PC1     & $0.293$      & $0.025$      & $0.083$      & $-1872.133$ & $0.104$ \\
\end{tabular}
\end{ruledtabular}
\end{table}

In order to examine the dependence of our results on the functional form and
on the effective interaction, we also studied $^{246}$Cm,
$^{248}$Cf, $^{250}$Fm, and $^{252}$No with other covariant density
functionals and several other typical parameter sets,
including PC-PK1~\cite{Zhao2010_PRC82-054319},
DD-ME1~\cite{Niksic2002_PRC66-024306}, and DD-ME2~\cite{Lalazissis2005_PRC71-024312}.
The results are listed in Table~\ref{tab:parameter}.

Roughly speaking, the results are similar with different parameter sets:
$\beta_{22}$ and $\beta_{30}$ vanish in all cases and for a specific nucleus,
the calculated $\beta_{20}$ and $\beta_{40}$ and the binding energy $E$ with
different parametrizations agree with each other.
For $^{246}$Cm, three parameter sets PC-PK1, DD-ME1, and DD-ME2 predict
zero $\beta_{32}$. With DD-PC1, as discussed before, a rather shallow
potential energy curve with a minimum at $\beta_{32}=0.020$ is obtained.
This indicates that the octupole correlation in $^{246}$Cm is very weak.
Both from the value of $\beta_{32}$ and from the energy gain $E_\mathrm{depth}$
due to the $\beta_{32}$ distortion,
it is found that the density-dependent functionals gives stronger $Y_{32}$
effects than the functional with nonlinear self coupling does.
Among the density-dependent functionals, the meson exchange nucleon
interaction with the parameter set DD-ME2 gives the largest energy gain
$E_\mathrm{depth}$ for $^{248}$Cf, $^{250}$Fm, and $^{252}$No.
The evolution of the non-axial octupole $\beta_{32}$ effects along
the $N=150$ isotonic chain is almost independent of the form of energy
density functional and the parameter set: The $Y_{32}$ effects is
the strongest in $^{248}$Cf except for that DD-ME1 gives a smaller energy
gain for $^{248}$Cf than for $^{250}$Fm.

In summary, we studied the non-axial octupole $Y_{32}$ effects in
$N=150$ isotones $^{246}$Cm, $^{248}$Cf, $^{250}$Fm, and $^{252}$No
by using multi-dimensional constraint covariant density functional theories.
Due to the interaction between a pair of neutron orbitals,
$[734]9/2$ originating from $\nu j_{15/2}$ and
$[622]5/2$ originating from $\nu g_{9/2}$, and
that of a pair of proton orbitals,
$[521]3/2$ originating from $\pi f_{7/2}$ and
$[633]7/2$ originating from $\pi i_{13/2}$,
rather strong non-axial octupole $Y_{32}$ effects have been found for $^{248}$Cf and
$^{250}$Fm which are both well deformed with large axial-quadrupole deformations,
$\beta_{20} \approx 0.3$.
For $^{246}$Cm and $^{252}$No, a shallow minima develops along
the $\beta_{32}$ deformation degree of freedom.
The evolution of the non-axial octupole $\beta_{32}$ effect along
the $N=150$ isotonic chain is not very sensitive to the form of
the energy density functional and the parameter set we used.
Finally we note that a nucleus with a very flat potential energy curve
$E \sim \beta_{32}$ may not have a static non-axial octupole $\beta_{32}$ deformation.
The ground state wave function should
be a strongly correlated superposition of different shapes with similar energies.
This suggests that further
investigations of these nuclei should include beyond mean
field effects by using, for example, the generator coordinate
method~\cite{Zberecki2009_PRC79-014319, Yao2010_PRC81-044311}.

Helpful discussions with Y. S. Chen, R. Jolos, Lulu Li, A. Lopez-Martens, Zhen-Hua Zhang,
and Kai Wen are gratefully acknowledged.
This work has been supported by
973 Program of China (2013CB834400),
NSF of China (10975100, 10979066, 11121403,
11175252, 11120101005, and 11275248),
and
Chinese Academy of Sciences (KJCX2-EW-N01 and KJCX2-YW-N32).
The results described in this paper are obtained on the ScGrid of
Supercomputing Center, Computer Network Information Center of Chinese Academy
of Sciences.


%

\end{document}